\documentclass[a4paper,review,fleqn]{cas-sc}

\usepackage{braket}
\usepackage{tikz}
    \usetikzlibrary{quantikz}
\usepackage{algorithm}
\usepackage{appendix}

\usepackage{algpseudocode}
\usepackage[normalem]{ulem}

\usepackage{diagbox}
\usepackage[numbers]{natbib}

\newcommand{\qop}[2]{\hat{#1}_{\mathrm{#2}}}
\newcommand{\qstate}[1]{\ket{\psi}_{\mathrm{#1}}}
\newcommand{\mathsc}[1]{{\normalfont\textsc{#1}}}

\makeatletter
\newcommand*\dashline{\rotatebox[origin=c]{90}{\ensuremath{\dabar@\dabar@}}}
\makeatother
\newcommand{\coloredtext}[2][black]{\textcolor{#1}{#2}}

\begin{document}

\def\tsc#1{\csdef{#1}{\textsc{\lowercase{#1}}\xspace}}
\tsc{WGM}
\tsc{QE}
\tsc{EP}
\tsc{PMS}
\tsc{BEC}
\tsc{DE}


\let\WriteBookmarks\relax
\def\floatpagepagefraction{1}
\def\textpagefraction{.001}

\shorttitle{QLGA}

\shortauthors{Bastida et~al.}

\title [mode = title]{Efficient Quantum Lattice Gas Automata}                      



%
\author[1,2]{Antonio David Bastida Zamora}[]

\affiliation[1]{organization={Quanscient Oy},
    addressline={Åkerlundinkatu 8}, 
    city={Tampere},
    postcode={33100}, 
    country={Finland}}

\affiliation[2]{organization={Aalto University},
    addressline={Otakaari 24}, 
    city={Espoo},
    postcode={02150}, 
    country={Finland}}

\credit{Data curation, methodology, software, validation, formal analysis, investigation, visualization, writing -- original draft}

\author[1,3]{Ljubomir Budinski}[]

\affiliation[3]{organization={Faculty of Technical Sciences, University of Novi Sad},
    addressline={Trg Dositeja Obradovića 6}, 
    city={Novi Sad},
    postcode={21000}, 
    country={Serbia}}

\credit{Conceptualization, methodology, software, formal analysis, investigation, resources,  visualization, writing -- original draft}

\author[1]{Ossi Niemim\"aki}[orcid=0000-0002-6860-0542]

\credit{Writing -- review \& editing}



\author
[1]
{Valtteri Lahtinen}

\credit{Supervision, funding acquisition, project administration, writing -- review \& editing}





\begin{abstract}
This study presents a novel quantum algorithm for lattice gas automata simulation with a single time step, demonstrating logarithmic complexity in terms of $CX$ gates. The algorithm is composed of three main steps: collision, mapping, and propagation. A computational complexity analysis and a comparison using different error rates and number of shots are provided. Despite the impact of noise, our findings indicate that accurate simulations could be achieved already on current noisy devices. This suggests potential for efficient simulation of classical fluid dynamics using quantum lattice gas automata, conditional on advancements to expand the current method to multiple time steps and state preparation.
\end{abstract}


\begin{highlights}
\item Efficient quantum lattice gas automata (QLGA) in the number of $CX$ gates
\item New methodology for simulating fluid flows using QLGA
\item Higher resilience to noise compared to similar algorithms
\end{highlights}


\maketitle

\section{Introduction}
\label{sect:Chapter 1}

Cellular automata have been widely explored in computational physics due to their ability to transform simple microdynamical rules into a specific macroscopic behavior. A type of cellular automaton called lattice gas automaton (LGA) was devised to simulate fluid dynamics effectively. The first LGA models to appear in the literature were the Hardy-Pomeau-Pazzis (HPP)~\cite{Hardy_Pomeau_1972,Hardy_Pazzis_Pomeau_1976} and Frisch-Hasslacher-Pomeau~\cite{Frisch_Hasslacher_Pomeau_1986} models. However, the study of LGA also lead to lattice Boltzmann methods (LBM)~\cite{Chen_Doolen_1998, Succi_LBM}, and LGA was largely abandoned due to the advantages presented by the LBM, specifically the noise resilience (in contrast to the local fluctuations in the LGA) and flexibility in simulating complex multiphysics. However, with the surge of quantum computing, LGA could be one of the key methods to realize the quantum advantage, as it  provides a simple framework for encoding lattice points and can model fluid evolution even in nonlinear models. Specifically, LGA has been proven to be able to simulate the Navier-Stokes equations using the Frisch-Hasslacher-Pomeau model, a more complex version of the preceding HPP model.

In recent years we have seen several attempts at bringing the LGA and LBM methods to quantum computing. The first attempt to quantize cellular automata and lattice gas automata models was made by Meyer~\cite{Meyer_1996}. Insisting on the exact unitarity of the collision operator to maintain consistency with standard quantum mechanics, the author formulated a collision operator for the simplest one-dimensional quantum cellular automaton and quantum LGA (QLGA) models. Later this single particle QLGA was used to simulate the Schrödinger equation~\cite{Meyer_1997,Boghosian_Taylor_1998}. To extend the applicability of the QLGA beyond the simulation of quantum systems, in particular to computational fluid dynamics, Yepez~\cite{Yepez_2001} derived a lattice Boltzmann equation that exactly describes the kinetic transport at the mesoscopic scale in a quantum lattice gas. In other words, he showed that the lattice Boltzmann equation is an exact representation of the particle dynamics, including all the effects due to quantum superposition and entanglement. This then lead to QLGA algorithms for the simulation of the diffusion equation~\cite{YEPEZ_2001_diff} and Burgers equation~\cite{YEPEZ_2002} on a one-dimensional lattice with two qubits per lattice site on a type-II quantum computer.%
~\footnote{Term used by Yepez to distinguish globally phase-coherent quantum computers (Type I) from local phase-coherent quantum computers (Type II) }

Although this QLGA model makes the simulation of classical physics possible on quantum computers, several issues need to be addressed before it can provide a more efficient alternative to the classical LGA. 
Namely, the requirement of one qubit per lattice site, computation of the propagation step classically, and measurements after each time step. One of the recent attempts to address the time concatenation and the propagation step was made in \cite{chrit2023fully}. In this model the propagation follows from a permutation block of swap gates, while the need for measurement until the end of computation is delayed by approximating the qubit relative phases and subtracting them at the end of each time step. However, the major limitation of using one qubit per lattice site still remains.

What then makes the QLGA so appealing for simulating fluid flows (or some other physical processes) on quantum devices? There is a variety of different methods for solving the fluid flow equations on quantum devices already available. Linearized differential equations can be solved with the linear system solvers~\cite{Liu_2021,Krovi_2023} and variational algorithms~\cite{Lubasch_2020,jaksch2022variational} and one can even apply several lattice-Boltzmann-based models~\cite{Budinski1,Budinski2,Blaga}. However, all of these methods suffer from the same illness in that they are not suitable for handling nonlinearity. Since in LGA nonlinearities emerge from particle interactions, this approach is an excellent candidate for simulating nonlinear multiphysics on quantum computers.

In this paper we present a new method for simulating fluid flows using the lattice gas automata models on quantum devices. Particularly, we address the issue of lattice representation from the previous quantum LGA algorithms (one qubit per lattice site) by exploiting the superposition property of all lattice sites. This feature allows us now to design a set of quantum algorithms that scale logarithmically with the number of states, exhibiting therefore an exponential speedup for one time step. We demonstrate this with one-dimensional LGA and two-dimensional HPP models. In the case of a simulation with multiple time steps, the algorithm requires measurement and initialization, as it is impossible to realize more than one iteration directly with the proposed quantum circuit.  Additionally, these models seem to be more resilient to quantum noise than the closely-related quantum lattice Boltzmann method, which is very sensitive to changes in the state amplitudes. LGA also solves the issue of simulating nonlinearities, as they are introduced as a result of local interactions between particles, exhibiting a complex behavior \cite{Bhattacharjee_Naskar_Roy_Das_2018}. The work focuses on simulations of the D1Q3 (advection-diffusion in 1D using three qubit channels) and HPP models.

\section{Lattice gas automaton}
\label{sect:Chapter 2}

LGA is a cellular automaton that describes the evolution of a fluid~\cite[Ch.~2]{Chopard_Droz_1998} \cite[Ch.~2]{Rivet_Boon_2005}. It is composed of the following elements:
\begin{enumerate}
    \item A regular lattice of cells, or lattice sites covering a $k$-dimensional space. The characteristics of the latticewill play a major role in the modelling capacity of the automaton. Specific lattice symmetries are often necessary for the simulation of partial differential equations. In what follows we assume periodic lattice boundaries.

    \item A set of Boolean variables, called channels, $n(r,t)=\{n_1(r,t),n_2(r,t),n_3(r,t),...,n_C(r,t)\}$ such that $n\in C$ with $C$ the set of all channels, $r \in \mathbb{R}^k$ and $t$ indicates a given time step. The channels provide information on the occupation ($n_i=1$) or lack ($n_i=0$) of particles defined at each time step $t$ and lattice site $r$. 
    
    \item A microdynamical equation  $n_i(r+v_i,t+1)=n_i(r,t) + \Delta_i[n(r,t)]$, which describes the time evolution of particles moving from a lattice site situated at position $r$ to the neighboring site located at position $r+v_i$ with the prescribed velocity $v_i$. The function $\Delta_i: C\rightarrow C$ is called the collision operator~\cite[Ch.~2]{Rivet_Boon_2005}. 

    \item Derivation of the physical variables is managed by introducing the averaged value of the channel occupation $\braket{n_i}$, where the mass density is defined as $\rho=\sum_i\braket{n_i}$. In contrast, the momentum density is calculated as $\rho u_{\alpha} = \sum_i \braket{n_i}v_{i\alpha}$. Average is taken over an ensemble of experiments with different initial conditions or some small sub-region. 
\end{enumerate}

We note that the collision rule $\Delta_i$ is applied to every lattice site at the same time, obeying the conditions of conserving mass $\sum_i \Delta_i(n) = 0$  and momentum $\sum_i  v_i \Delta_i(n) = 0$. This means that the time evolution is synchronous for each lattice site. At the same time, the dynamics are homogeneous, meaning that the collision rules must be equal in every lattice site.

\subsection{The D1Q3 model: collision rules and microdynamics}
Several different lattice gas models can be constructed depending on the spatial dimensions and the desired physics. Here we will introduce the 1D and 2D models which simulate the propagation and diffusion of a fluid in the domain.

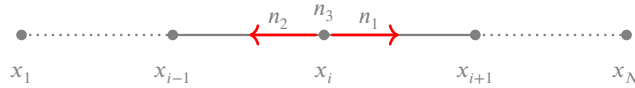
\begin{figure*}[ht]
    \centering
    \usetikzlibrary{arrows.meta}

\tikzset{
  myarrow/.style={
    line width=1pt,
  }
}

\begin{tikzpicture}[scale=1, transform shape]
    \fill (-2,0) circle (2pt) node[below=3mm] {$x_{1}$};
    \fill (0,0) circle (2pt) node[below=3mm] {$x_{i-1}$};
    \fill (2,0) circle (2pt) node[below=3mm] {$x_{i}$};
    \fill (4,0) circle (2pt) node[below=3mm] {$x_{i+1}$};
    \fill (6,0) circle (2pt) node[below=3mm] {$x_{N}$};
    \draw[dotted,line width=0.3mm] (-2,0)--(0,0);
    \draw[line width=0.3mm] (0,0)--(2,0);
    \draw[line width=0.3mm] (2,0)--(4,0);
    \draw[dotted,line width=0.3mm] (4,0)--(6,0);

    \draw[line width=0.3mm]  (1.4,-0.05)  node[above=0.1mm] { $n_2$};
    \draw[line width=0.3mm]  (2.00,0.05)  node[above=0.1mm] { $n_3$};
    \draw[line width=0.3mm] (2.6,-0.05)  node[above=0.1mm] { $n_1$};

    \draw[->,myarrow,red, shorten >=0.1mm] (2.1,0)--(3,0);
    \draw[->, myarrow,red,shorten >=0.1mm] (1.9,0)--(1,0);
\end{tikzpicture}
    \caption{D1Q3 lattice configuration.}
    \label{fig:lattice-d1q3}
\end{figure*}

\label{sect:D1Q3 Model}
The D1Q3 model for quantum computing applications was first defined as a hybrid method between LGA and LBM~\cite{Yepez_2001}. The 1D model was thought originally to have two different channels, the first with a particle going to the left and the second with a particle going to the right, both particles having a mass $m=1$. However, this leads to a very noisy probabilistic profile with drastically different values on even and odd lattice sites. To avoid this problem, one can introduce an additional channel with a rest particle of mass $m=2$, adding diffusion to the dynamics (see Figure~\ref{fig:lattice-d1q3}). Therefore, we will set the Boolean model to be a binary string followed by three channel bits, corresponding to eight possible collision states. This means that each binary string consists of two registers, one associated with the channels, with one bit for each of them and another for the lattice site.
    
The collision rule is based on three channels and includes exchange between only two states. With the notation of $(n_1 n_2 n_3)$ for the encoding of the right, left and rest channels, respectively, we can use a lookup table depicted in Table~\ref{table_d1q3} to specify the collision rules for the D1Q3 model. When two particles collide on a given lattice site with opposite directions defined by a state $(110)$, a rest particle is formed giving the state $(001)$. In contrast, when we have only a rest particle as in state $(001)$, the result state $(110)$ is two particles moving in opposite directions. All other states remain unaffected by the collision process, ensuring that the mass and momentum are conserved.  After the collision step, the particle residing on the channel $n_1$ propagates one step to the right, whereas the particle on the channel $n_2$ is propagates one step to the left. The rest particle $n_3$ remains at the initial position. 
    
\begin{table}[!htbp]
\centering
    \caption{D1Q3 Model States.}
    \begin{tabular}{c c c}
    \hline
    \rowcolor{gray!50}
    \textbf{Initial State} & & \textbf{Final State} \\
    000 & $\longrightarrow$ & 000 \\
    001 & $\longrightarrow$ & 110 \\
    010 & $\longrightarrow$ & 010 \\
    011 & $\longrightarrow$ & 011 \\
    100 & $\longrightarrow$ & 100 \\
    101 & $\longrightarrow$ & 101 \\
    110 & $\longrightarrow$ & 001 \\
    111 & $\longrightarrow$ & 111 \\
    \hline
    \end{tabular}
    \label{table_d1q3}
\end{table}

In simulations and model interpretation, the physics is not captured by the microscopic evolution of the system but instead by the macroscopic behavior that emerges from the local interaction and propagation of particles. To this end we introduce an average value of the channel occupation number $\braket{n_i}$. A common practice is to take an average over an ensemble of experiments with different initial conditions or to perform spatial averaging using the sub-lattices. These methods allow us to calculate the average occupancy of a lattice site on the macroscopic scale reducing the noise.

\section{Quantum lattice gas automaton}
\label{sect:Chapter 3}

Quantum computing has the potential to provide a computational advantage by exploiting quantum phenomena like superposition and entanglement. For a classical model to utilize the quantum properties efficiently, the structure of the model needs to coincide or have some overlap with the mechanism of the evolution of a quantum system. In LGA, we can design the collision and the propagation operators in such a way that they can exploit quantum properties. To demonstrate the LGA as a quantum-native method for simulating fluid flow, we present a set of quantum algorithms for simulations on quantum devices. Particularly, we focus on quantum algorithms for one-dimensional LGA that use the quantum superposition property to encode the lattice structure. In these algorithms, the collision is local and, therefore, can be addressed in superposition, while the propagation is not. However, the locality is reestablished from the perspective of each particle's channel using an intermediary mapping step, which changes the system's encoding. Therefore, a logarithmic scaling of the lattice sites can be achieved. First, by using a unitary operator for the collision and then by changing the encoding to allow the shifting of the particles using a unitary operator, depending on the channel of each particle. However, this method does not allow the change of the encoding back from the superposition to the binary, as the superposed states are not orthogonal to each other and cannot be perfectly distinguished. This problem was already discussed by Schalkers \cite{schalkers2024importance} and avoiding this is a current research topic.   

\subsection{D1Q3 quantum model}
The quantum LGA algorithm mostly follows the structure of the classical LGA and can be defined in five steps: initialization, collision, propagation, measurements, and post-processing. Each of these steps requires a particular algorithm design tailored to quantum devices. The superposition principle for the lattice representation in LGA provides a certain level of flexibility for building the initialization and collision steps. As a direct consequence, we have developed two different solutions for the encoding and collision. The first approach uses the binary representation of the channels,  as in the classical LGA. In contrast, the second approach has an encoding step that uses the superposition of possible states. Notice that in both algorithms, the number of qubits used scale as $O(\log_2(N))$, with $N$ the number of lattice sites. In what follows, we provide a detailed look into the QLGA algorithms for each method, complexity analysis, and results from simulations with depolarizing error rates based on real-devices.
\begin{figure}
	\centering
	\input{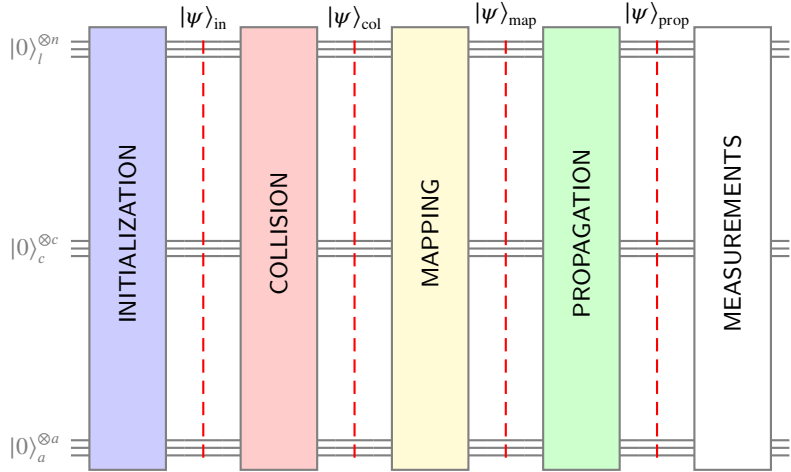}
    \caption{Quantum circuit for the QLGA.}
    \label{fig:circuit-qlga}
\end{figure}

\subsubsection{Binary-based quantum LGA algorithm}

Let us divide the binary-based quantum LGA into the first four steps mentioned above: initialization, collision, propagation, and measurements. In addition to these, we will also need another step called \emph{superposition mapping} in between the collision and propagation in order to use the state-of-the-art quantum algorithms for the propagation step.

\paragraph{Initialization} The initialization is one of the most important steps of any quantum algorithm with complex initial conditions. Initializing non-structured data can lead to the number of multi-controlled gates growing exponentially with respect to the number of qubits. Several methods have been researched to reduce the complexity of the initialization \cite{Mottonen_Vartiainen_Bergholm_Salomaa_2005, Araujo_Park_Petruccione_da_Silva_2021,Mozafari_Riener_Soeken_De_Micheli_2021}. However, for purely practical reasons we adopt the reverse iterative procedure proposed by Shende et al. \cite{Shende_2006} and focus on the implementation of the LGA itself. As mentioned by Shende, the state preparation consists of no more than $\frac{23}{48} 4^n - \frac{3}{2}2^n+\frac{4}{3}$ $CX$ gates, where $n$ are the number of qubits. Therefore, the state preparation scales exponentially and presents an obstacle to obtain quantum advantage, as is the case for a majority of current algorithms for quantum devices. As mentioned, a lower amount of $CX$ gates can be obtained, depending on the initial condition imposed. However, the exponential scaling with the number of qubits encoding the lattice sites remains.

Three channels per lattice site are required to encode the initial channel occupancy for the D1Q3 LGA model as in Section~\ref{sect:Chapter 2}. To encode this we introduce a three-qubit register $c$. Each qubit in this register stores the occupancy of a specific channel: right, left, and rest. Furthermore, we encode a one-dimensional lattice in the superposition of basis states of a quantum register $l$ with $n=\log_2(N)$ qubits, where $N$ is the number of sites in the lattice. Additionally, we need a three-qubit register $a$ for the switch from the binary representation of the channels to the superposition. This register is not needed for the encoding itself, but it later helps to implement the propagation using well-known state shift techniques~\cite{Budinski_2023, Shakeel_2020}.

Let us denote this encoding of the inital channel occupancy as an operator $\qop{I}{D1Q3}$, so that the initial quantum state will be as follows:
\begin{equation}
    \begin{split}
    \qstate{in}=\qop{I}{D1Q3} \qstate{0} & = \qop{I}{D1Q3} \left(\ket{0}_l^{\otimes n} \otimes \ket{0}_c^{\otimes 3}\right) \otimes \ket{0}_a^{\otimes 3}  \\ & =  \frac{1}{\sqrt{2^n}}\sum_{i=0}^{2^n-1} d_i \otimes \ket{0}_a^{\otimes 3} =
     \frac{1}{\sqrt{2^n}}\sum_{i=0}^{2^n-1} \ket{l_1....l_n |\overset{n_1 n_2 n_3}{c_1 c_2 c_3}}_i \otimes \ket{0}_{a1} \otimes \ket{0}_{a2} \otimes \ket{0}_{a3},
    \end{split}
\end{equation}
where $d_i$ are binary strings (with respect to the canonical basis states) composed of the $l$ bits encoding the lattice site position, and the $c$ bits storing a possible occupancy combination. For example, the binary string $\ket{0001 | 010}$ would refer to the occupancy of $\ket{010}$ (left channel active) at the lattice site positioned at $\ket{0001}$. Note that the contents of the active channel $\ket{c_1c_2c_3}_i$ varies depending on the position $i$. In all the equations that follow, we will sum over the lattice sites $i \in \{0, \dots, 2^{n-1}\}$: this is the \emph{ideal case} where no superposition of the channel states per lattice site is desired. In any realistic simulation, channel superpositions are however expected to appear due to the device noise. For full superposition states one needs to amend the indexing and renormalization to account for all eight possible channel states over each lattice site. However, this does not affect the structure of the algorithm, so we leave it out for simplicity.

\begin{figure}
    \centering
    \input{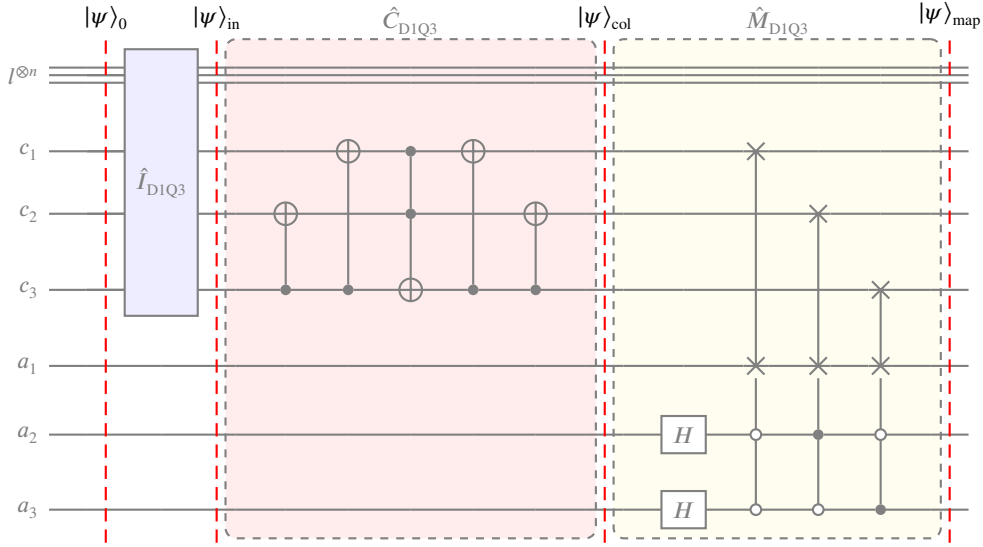}
    \caption{Quantum circuit for the initialization, collision, and mapping step.}
    \label{fig:init_coll_map}
\end{figure}

\paragraph{Collision} In the collision step we switch between two particular occupancy sub-states as in Table~\ref{table_d1q3}. Since the occupancy of the channels is stored only in the $c$ register, the collision step involves only those three qubits. Four $CX$ gates and one Toffoli gate are required to switch between occupancy sub-states $\ket{110}_c$ and $\ket{001}_c$ while leaving all other states unaffected, see Figure~\ref{fig:init_coll_map}. We will denote this gate sequence as the operator $\qop{C}{D1Q3}$, and obtain the new state $\qstate{col}$. In the following we will mark the collided channel by $\ket{c'_1c'_2c'_3}$.

\paragraph{Superposition mapping} Next we need to propagate the occupancy of each channel spatially according to the prescribed velocity of the particular LGA model. 
For the D1Q3 LGA model, the particles residing on the channel pointing to the right/left are propagated one step at a time in the right/left direction, while the rest particle remains still. The practical implementation of this movement is straightforward classically, but the quantum implementation is more involved. Here we wish to use the efficient state shift method as detailed in \cite{Budinski_2023}, and this requires a rearrangement of the qubit encoding. Particularly, to propagate each qubit state independently in the $c$ register, the occupancy sub-state needs to be decomposed. The idea is to map each qubit from the $c$ register into a superposition of states using register $a$ entangled with the lattice sites. We do this by first bringing the qubits $a_2$ and $a_3$  from the $a$ register into a superposition state using Hadamard gates followed by three multi-control swap gates between the $c$ and $a$ registers, as depicted in Figure~\ref{fig:init_coll_map}. This gate sequence produces a new state $\qstate{map}$:        
\begin{equation}
    \begin{split}
    \qstate{map}=\qop{M}{D1Q3} \qstate{col} & = \mathsc{MCSWAP}_{a-c} 
     \frac{1}{\sqrt{2^n}}\sum_i \ket{l_1....l_n|c'_1 c'_2 c'_3}_i \otimes \ket{0}_{a1} \otimes H \ket{0}_{a2} \otimes H \ket{0}_{a3} \\ 
     & = \frac{1}{2\sqrt{2^n}} \left( \sum_i \ket{l_1....l_n|a_1 c'_2 c'_3}_i \otimes \ket{c'_1} \otimes \ket{00}_{a2,a3} \right.\\  
     & + \sum_i \ket{l_1....l_n|c'_1 a_1 c'_3}_i \otimes \ket{c'_2} \otimes \ket{10}_{a2,a3} \\ 
     & + \sum_i \ket{l_1....l_n|c'_1 c'_2 a_1}_i \otimes \ket{c'_3} \otimes \ket{01}_{a2,a3} \\ 
     & + \left. \sum_i \ket{l_1....l_n|c'_1 c'_2 c'_3}_i \otimes \ket{a_1} \otimes \ket{11}_{a2,a3} \right),
    \end{split}
\end{equation}
where we now have mapped each channel of the register $c$ into the sub-states marked with the corresponding state of qubits $a_2$ and $a_3$. With this we obtain an equal superposition of the lattice sites (register $l$) and each channel (register $c$), which now allows us to apply one-step shift in different directions independently for each channel.

\begin{figure}
	\centering
	\input{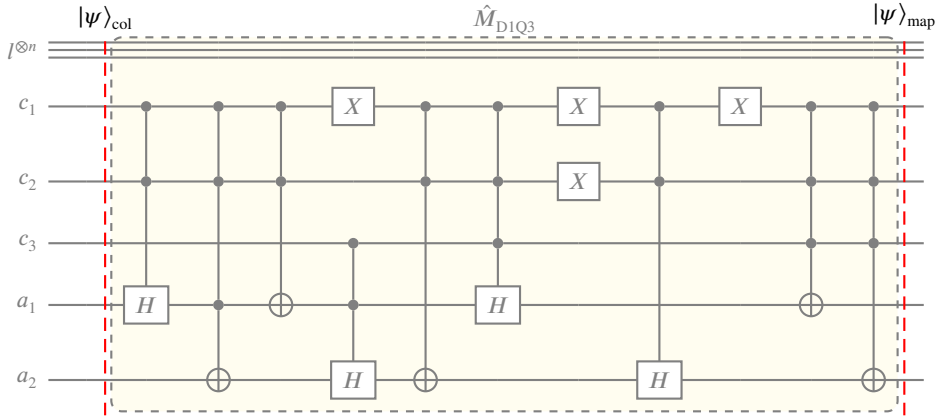}
 \caption{Quantum circuit for the mapping step without swap operations.}
 \label{fig:map_noswap}
\end{figure}
Alternatively, the mapping step can be performed using a more naive approach, where the information from the channels is translated into the ancilla register using Hadamard gates, resulting in a non-even superposition (not all the states have the same probability) between desired states (see Figure~\ref{fig:map_noswap}). As a result, we gain a junk-free state with a smaller number of ancilla qubits. The computational complexity of this approach is however significant in contrast to the first method, therefore we will not use this in the following.

\paragraph{Propagation} The propagation of the particles along the corresponding directions is implemented with the increment and decrement operations on the basis states. In the case of the D1Q3 LGA, the increment moves the particle $n_1$ one step to the right, while the decrement moves the particle $n_2$ one step to the left (see Figure~\ref{fig:lattice-d1q3}). Particle $n_3$ is at rest, therefore not changing its position during the propagation step.

\begin{figure}
	\centering
	\input{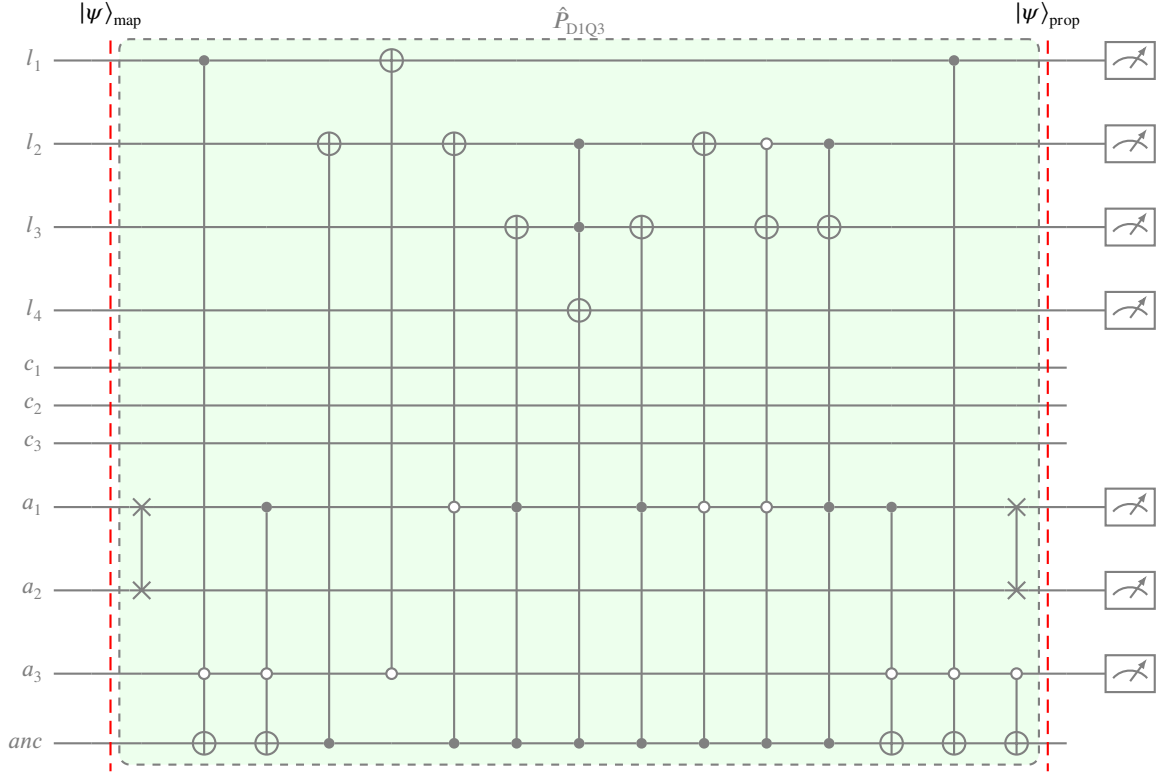}
 \caption{Quantum circuit for the propagation step.}
 \label{fig:prop}
\end{figure}

Several quantum algorithms exist for these shift operators, as they are crucial in several applications, such as the quantum random walk~\cite{Childs2}, matrix block encoding~\cite{Daan}, and the quantum lattice Boltzmann method~\cite{Budinski1,Budinski2,Blaga}. For efficient implementation, one could use a shift based on the quantum Fourier transformation proposed by Shakeel~\cite{Shakeel_2020}, or a parallel shift method by Budinski et al.~\cite{Budinski_2023}. An implementation of the parallel shift for the D1Q3 QLGA in the case of $16$ lattice sites (four qubits in the $l$ register) is shown in Figure~\ref{fig:prop}, with an additional one-qubit ancilla register. Expansion of the parallel shift to the larger number of lattice sites is straightforward~\cite{Budinski_2023}. Applying the propagation operator $\qop{P}{D1Q3}$ on state $ \qstate{map}$ results in a new state:
\begin{equation}
    \begin{split}
    \qstate{prop}=\qop{P}{D1Q3} \qstate{map} & = \qop{P}{D1Q3} \frac{1}{2\sqrt{2}} \left( \sum_i \ket{l_1....l_n|a_1 c'_2 c'_3}_i \otimes \ket{c'_1} \otimes \ket{00}_{a2,a3} \right.\\  
     & + \sum_i \ket{l_1....l_n|c'_1 a_1 c'_3}_i \otimes \ket{c'_2} \otimes \ket{10}_{a2,a3} \\ 
     & + \sum_i \ket{l_1....l_n|c'_1 c'_2 a_1}_i \otimes \ket{c'_3} \otimes \ket{01}_{a2,a3} \\ 
     & + \left. \sum_i \ket{l_1....l_n|c'_1 c'_2 c'_3}_i \otimes \ket{a_1} \otimes \ket{11}_{a2,a3} \right)\\
     & = \frac{1}{2\sqrt{2}} \left( \sum_i \ket{l_1....l_n}_i \otimes \ket{a_1 c'_2 c'_3}_{i+1} \otimes \ket{c'_1} \otimes \ket{00}_{a2,a3} \right.\\  
     & + \sum_i \ket{l_1....l_n}_i \otimes \ket{c'_1 a_1 c'_3}_{i-1} \otimes \ket{c'_2} \otimes \ket{10}_{a2,a3} \\ 
     & + \sum_i \ket{l_1....l_n}_i \otimes \ket{c'_1 c'_2 a_1}_i \otimes \ket{c'_3} \otimes \ket{01}_{a2,a3} \\ 
     & + \left. \sum_i \ket{l_1....l_n}_i \otimes \ket{c'_1 c'_2 c'_3}_i \otimes \ket{a_1} \otimes \ket{11}_{a2,a3} \right).
    \end{split}
\end{equation}
Note that now the binary strings $d_i$ representing the position and the channels have to split, as we move the channel information along the directions on the lattice. The lattice indexing is assumed to be periodic. With this propagation step, we complete one time step of the D1Q3 LGA model. 

\paragraph{Measurements} The registers $l$ and $a$ are measured at the end of each time step. As the encoding in this model is binary, the states do not carry any information in the amplitudes or the phases. Information about the occupancy of the channel is stored in the qubit $a_1$, and the particular channel is identified with the states of qubits $a_2$ and $a_3$. In other words, when the state $\ket{00}_{a2,a3}$ is observed, the qubit $a_1$ corresponds to the right moving channel, while an observation of the state $\ket{10}_{a2,a3}$ means that the qubit $a_1$ denotes the left moving channel. Similarly, the state $\ket{01}_{a2,a3}$ marks the resting particle, while the remaining state $\ket{11}_{a2,a3}$ is a computational junk state. We then use these measurement results to prepare the initial state for the next time step.

\paragraph{\coloredtext{Simulating multiple time steps}}

\coloredtext{The current algorithm only simulates one time step.} Therefore measurement is required in this version of the algorithm as there does not exist a unitary operator $U$, such that $ \alpha_1 \ket{001}_l + \alpha_2\ket{010}_l + \alpha_3\ket{100}_l \rightarrow \ket{\alpha_1 \alpha_2 \alpha_3}_l $, where $\alpha_i = 0,1$ for $i={1,2,3}$, 
neglecting the normalization factor. This is because the superposition of the channels cannot be recombined uniquely due to non-orthogonality of the states. 

\subsubsection{Superposition-based quantum LGA algorithm}
The superposition-based variation avoids the intermediary mapping by encoding the channels differently. Propagation requires a superposition of not only the lattice sites but also channels, suggesting one should use a consistent encoding throughout each step. In order to use the encoding corresponding to the superposition of lattice sites and particles from the beginning of the algorithm, we need to be able to realize the collision in a complete superposition. However, this is not possible, as each state has information about one particle and its velocity but not about the other particles occupying each site. This presents a challenge for the collision, as it is impossible to use the lookup table without knowing which particles are occupied. As we noted in the introduction of this section, this issue has been previously discussed already by Schalkers \cite{schalkers2024importance}. Nevertheless, this issue can be addressed by incorporating additional operations as a classical post-processing step. This means that most of the collision operator is prepared already during the post-processing, making this alternative a hybrid approach. Given that every state must be, in any case, iterated over classically after the measurements to prepare for the next initialization, we find this alternative useful to perform simulations and test the results with a lower computational requirement. 

\paragraph{Initialization} For the encoding part only occupied channels are taken into account. The channel register $c$ is again composed of three qubits, but instead of marking each channel with a particular qubit, we will encode them in a superposition. Particularly, the rest particle is encoded in the state \coloredtext{$\ket{001}_c=\ket{1}_c$}, occupied left channel is mapped to the state $\ket{010}_c=\ket{2}_c$, finally the right channel is encoded in the state \coloredtext{$\ket{100}_c=\ket{4}_c$}. In the case of the D1Q3 LGA model, only two possible states are affected by the collision operator, that is, the state that denotes the presence of only rest particle $\ket{001_c}$ and the superposition of the left and right particles $\ket{010}_c + \ket{100}_c$. To apply the collision only on those two states while leaving all others fixed, we introduce an additional swapping in the form $\ket{001}_c \mapsto \ket{111}_c$ and $\left(\ket{010}_c + \ket{100}_c\right) \mapsto \ket{000}_c$. After applying the initialization operator $\qop{I}{D1Q3}$ we obtain a new state that encodes the initial occupancy:

\begin{equation}
    \begin{split}
      \qstate{in}=\qop{I}{D1Q3} \qstate{0}    &  = \qop{I}{D1Q3} \left(\ket{0}_l^{\otimes n} \otimes \ket{0}_c^{\otimes 3}\right) \otimes \ket{0}_a^{\otimes 3}  = \frac{1}{\sqrt{2^n}}\sum_{i=0}^{N-1} \ket{i}_l \otimes \sum_{k=1,2,4} s_k \ket{k}_c \otimes \ket{0}_{a}^{\otimes3} ,
    \end{split}
\end{equation}

where the parameter $s$ can take values $0$ or $1$ as defined by the following conditions:

\begin{equation}
    \begin{cases}
       \ket{1}_c \mapsto \ket{7}_c , & s_4=0, s_2=0, s_1=1  \\
       \ket{2}_c+\ket{4}_c \mapsto \ket{0}_c ,  & s_4=1, s_2=1, s_1=0 .
    \end{cases}
\end{equation}

\begin{figure}
	\centering
	\input{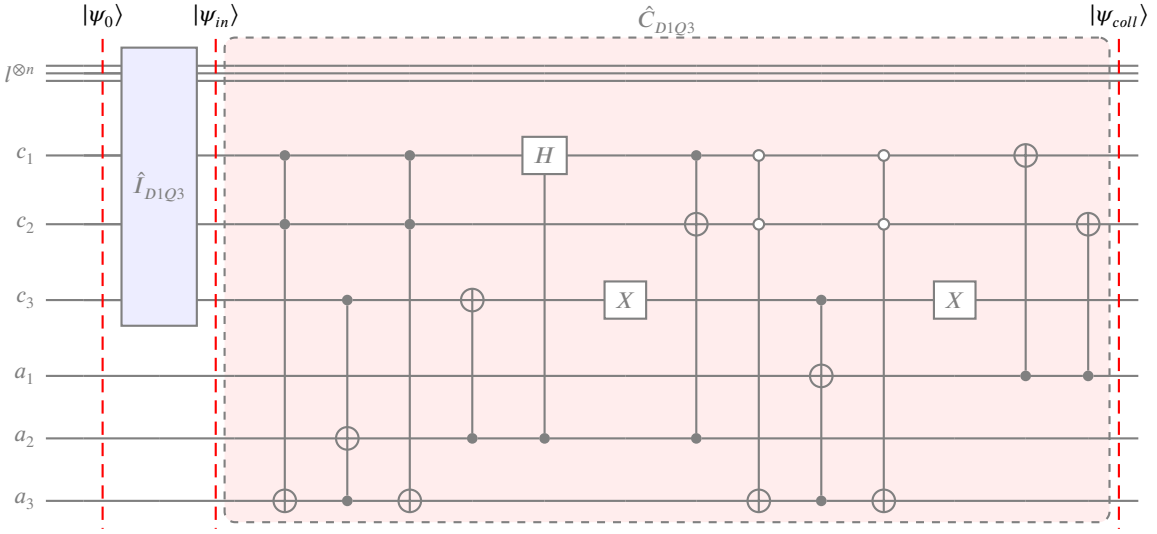}
 \caption{Quantum circuit for the initialization and collision step in the superposition variant.}
 \label{fig:init_coll_super}
\end{figure}

\paragraph{Collision} During the collision, we transform the state encoding the presence of left and right particles $\ket{000}_c$ into $\ket{001}_c$ (rest only), while the state $\ket{111}_c$ is cast into the superposition of the left and right particles $\ket{010}_c + \ket{001}_c$. All other states are unaffected. The quantum circuit for the collision is shown in Figure~\ref{fig:init_coll_super}. While the collision rule applied is the same as in the previous variant of the algorithm, the quantum circuit applied is different. This is because in this case, we encode classically in which registers we have to do the collision and to which state. This transformation is later done as an unitary operator, applied to the states to be changed. In specific $\ket{000}_c$ and $\ket{111}_c$.

\paragraph{Propagation} One of the possible variants of the base shift described in the previous section can be used for the propagation step. The pseudo-code of the entire workflow is shown in Algorithm~\ref{alg:qlga_alt}.
\begin{algorithm}
    \caption{Superposition-based quantum LGA algorithm}
    \label{alg:qlga_alt}
    \begin{algorithmic}[1]
    \Procedure{QLGA}{}
        \State $\qstate{in} \gets$ Initialize $\qstate{0}$ 
        \For{$t = 1$ to $T$} \Comment{$T$ is the total number of time steps}
            \State $\qstate{col} \gets$ Collide $\qstate{in}$ \Comment{Collision affects only the states marked during post-processing}
            \State $\qstate{prop} \gets$ Propagate $\qstate{col}$
            \State $measurement \gets$ Measure $\qstate{prop}$
            \State $state \gets$ Post-process $measurement$ \Comment{Mark $\ket{111}$ if only rest particle in a lattice site; Mark $\ket{000}$ if only right and left channels are present}
        \EndFor
    \EndProcedure
    \end{algorithmic}
\end{algorithm}

\paragraph{\coloredtext{Simulating multiple time steps}}
In this case, we prepare the quantum state in a way that is easy to implement for the collision and propagation. However, measurement and reinitialization are necessary as the state vector needs to be arranged in a particular manner for the next collision. \coloredtext{Therefore, the current quantum algorithm only allows the simulation of a one time step, while multiple time steps are possible by a new initialization after extracting the data}. Effectively, this makes this alternative to be a hybrid approach that requires measurement at the end of each time step.

\section{Computational complexity}
\label{sect:complexity}

Let us next look at the gate complexity of the quantum LGA algorithms. The focus of this analysis will be on three main steps that make the core of the D1Q3 QLGA algorithm: collision, mapping, and propagation. Amplitude encoding is one of the major bottlenecks in several quantum algorithms using classical data encoding, and the initialization step could potentially be reduced, as commented before, in terms of the number of $CX$ gates. As in most current quantum algorithms, state preparation is an obstacle for quantum advantage, as it requires exponential number of $CX$ gates. While further optimizations could be accomplished, the exponential scaling remains. Therefore, initialization and measurement will not be considered in the following computational analysis. Notice that the following analysis is related to one time step, while multiple-time steps require measurement and initialization, both scaling as $O(N)$, with $N$ lattice sites. 

The collision and the mapping step between binary encoding to superposition scales as $O(1)$ as the number of channels is constant. On the other hand, the propagation depends critically on multi-controlled gates and will be the source for most of the complexity in the algorithm. We have implemented the propagation using the basis state shift: an extensive complexity analysis is provided in \cite{Budinski_2023} for the canonical shift, the QFT variation, and the parallel shift, together with an outline of an ancilla decomposition of the multi-controlled gates~\cite[pp.~183--184]{Nielsen_Chuang_2022}. The results focus on a transpilation with the IBM basis set of gates $CX$, $I$, $RZ$, $SX$ and $X$ using standard compilation techniques. It can be concluded that the best choice for conducting the shift step for larger lattices is the parallel shift algorithm, as it provides a linear scaling $n_{CX}(n) = 15(n-6) + 149$ in terms of $CX$ gates, when the number $n$ of working qubits is at least six. Note that this scaling requires to use an additional ancilla register for the proper decomposition of the multi-controlled gates into generalized Toffoli gates, many of which then cancel each other.

In summary, the algorithm has a logarithmic scaling of $CX$ gates in terms of the number of lattice sites. This is possible due to the binary encoding of the channels and the application of the same unitaries (collision and propagation) to every lattice site, which allows to simulate QLGA efficiently. 

In terms of space complexity, both algorithmic variants, make use of a logarithmic encoding of the lattice sites, as these are encoded in superposition from the beginning to the end. In specific, the D1Q3 algorithm uses $\log_2(N)+6$ qubits for both variants. The additional number in both algorithms comes from the channel and ancilla registers. If we expand this method to higher dimension, for example using the proposed extension to 2D HPP given in the Appendix~\ref{appendixA}, the number of qubits for the lattice register scales as $O(\log_2(NM))$, where variable $N$ represents the number of sites in the $x$-direction and variable $M$ the sites in the $y$-direction, supposing a rectangular lattice. However, this does not necessarily double the gates: for example, the one-dimensional propagation can be generalized to multi-dimensional lattices with the use of controlled swap gates, which add only a linear factor to the number of $CX$ gates needed.
\section{Numerical examples with the D1Q3 model}

To test the QLGA we carry out two benchmark tests: a comparison with the classical version, and an analysis on the effect of simulation noise. Due to the size of the system required for the simulation of a two-dimensional domain with the quantum HPP model, we focus on the D1Q3 QLGA.

\begin{figure}[hbt!]
    \centering
    \includegraphics[width=0.31\textwidth]{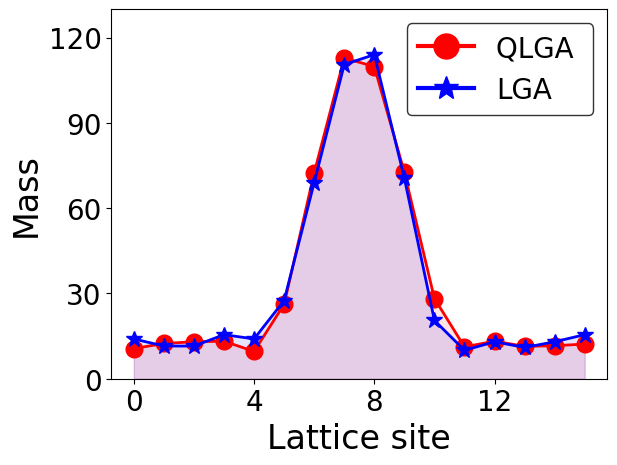}
    \includegraphics[width=0.31\textwidth]{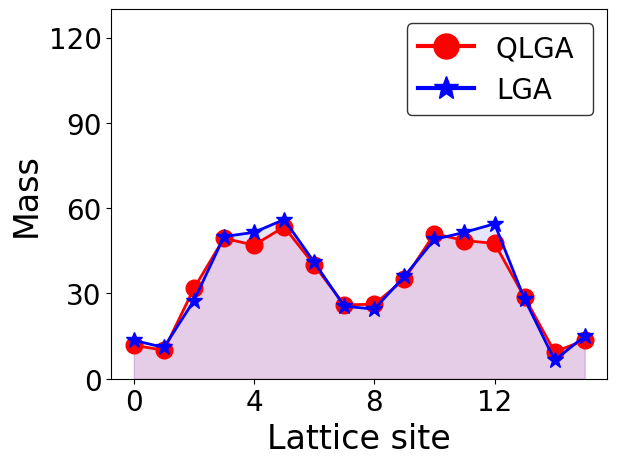}
    \includegraphics[width=0.31\textwidth]{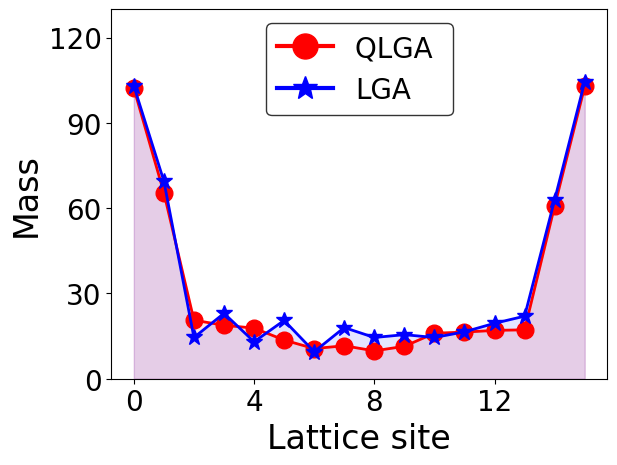}\\[\smallskipamount]
    \includegraphics[width=0.31\textwidth]{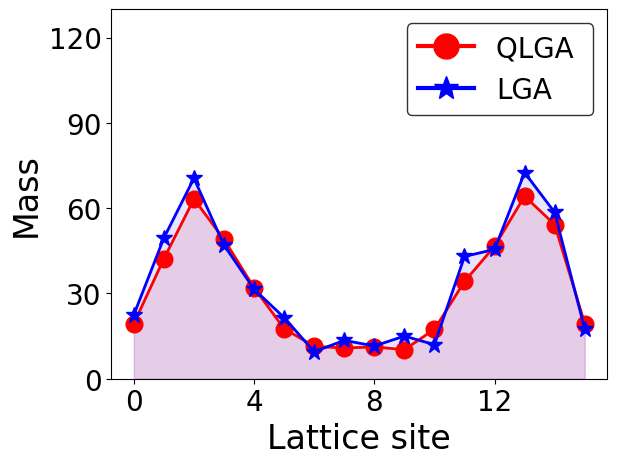}
    \includegraphics[width=0.31\textwidth]{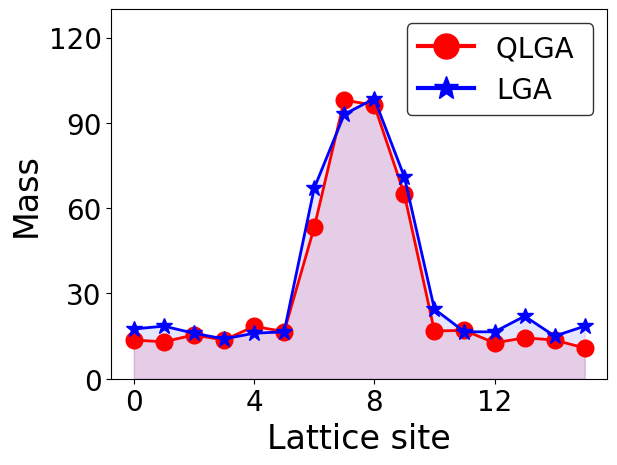}
      \includegraphics[width=0.31\textwidth]{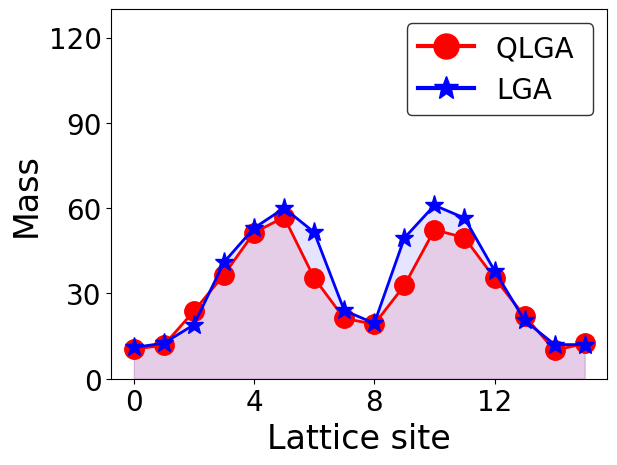}\\[\smallskipamount]
    \caption{Comparison between LGA and QLGA for different time steps ($t=0$, $t=160$, $t=320$, $t=560$, $t=752$, $t=896$). QLGA is shown in blue and the LGA version in red. A small difference is observed as the initial state is random, and the result is averaged over an ensemble.}
    \label{fig:comparison_clear}
\end{figure}

First, we compare the results of the quantum LGA against a classical version. For this we use a one-dimensional lattice with $512$ lattice sites, from witch $32$ sites where used as sub-lattices for spatial-averaging. This means that we output a mass value obtained as the sum of the $32$ sublattices that compose it. As each sublattice can have up to one rest particle ($m=2$) and left and right particles ($m=1$), the maximum total mass for each lattice site in the output is $m=128$. As the initial condition, we set up a sharp profile in the middle of the domain as $95~\%$ channel occupancy and $5~\%$ of probability for the rest of the model. We conducted several simulations using the different variants of QLGA, and validated against classical simulations, yielding the same results. One example of these experiments is illustrated in Figure~\ref{fig:comparison_clear}. The simulation was performed using the Aer simulator from Qiskit SDK~\cite{Qiskit} without noise. Besides the sub-lattice averaging, we have found that a statistical ensemble, with different initial conditions for each trial, is necessary to reduce the noise, due to the limited number of qubits in current simulators and quantum computers (more qubits are needed to increase the sub-lattice sites). The results clearly show a diffusion-like behavior for both of the classical LGA and QLGA models, and the results clearly overlap for each time step. Notice that the small differences that we observe during the simulation are a direct consequence of different initial states for $t=0$. This is because the initial condition is set randomly but with fixed probabilities and averaged over an ensemble of simulations to smooth out the final result. This is a common practice for LGA due to the noisy nature of the simulation. If the initial state is the same, the final result will also be identical and the more realizations of the simulation are done, the better the final result will be.
This validates that the presented quantum algorithms are capable of recreating the behavior of the classical LGA. 

Quantum computing is advancing fast and new devices and architectures with more qubits and less noise are continuously presented. However, noise is still unavoidable in quantum devices of today, and it is relevant to find algorithms that can be used under noisy conditions. Therefore, we also explored how the algorithm behaves under different error rates. Due to the high cost of quantum computers and limited access to machines with varying error rates, we tested the algorithm with the Qiskit Aer simulator and constructed simple models for the noise. The error rates of the machine were manually included. As shown in some works as \cite{PhysRevE.104.035309}, depolarizing errors are the major contributors to noise in real quantum devices. Other variables to consider are the number of shots to extract results, the connectivity of the machine, and the set of native gates used. To simplify the analysis as much as possible, we have selected the IBM native gate set, which is composed of $RZ$, $X$ , $SX$ and $CX$ gates, and tested the model under three noise levels. Each noise model has been created using single-qubit errors, two-qubit errors, a depolarizing error, and a readout error provided by Qiskit. In Table~\ref{table:error_rate}, we detail the different error rates used for the low, mid, and high noise conditions.
\begin{table}[hbt!]
    \centering
    \caption{Benchmarking Model Resilience to Noise}
    \begin{tabular}{|c|c|c|c|}
    \hline
    \textbf{Noise Level} & \textbf{Single Qubit Error Rate} & \textbf{2-Qubit Error Rate} & \textbf{Readout Error Rate} \\
    \hline
    Low & $10^{-5}$ & $10^{-4}$ & $10^{-4}$ \\
    Mid (similar to Quantinuum H2) \cite{Quantinuum2024} & $3 \times 10^{-5}$ & $2 \times 10^{-3}$ & $2 \times 10^{-3}$ \\
    High (similar to early quantum devices) & $6 \times 10^{-3}$ & $2 \times 10^{-2}$ & $2 \times 10^{-2}$ \\
    \hline
    \end{tabular}
    \label{table:error_rate}
\end{table}

Increasing the number of shots should lead to more accurate simulations. To test this assumption, we ran simulations for every noise level with the minimum number of shots required to observe propagation and diffusion up to time step $t=48$. In Table~\ref{table:comparison_noise} we can see how the simulation behaves. The higher the noise, the more shots are required to replicate similar results, as was expected. With increasing noise, it becomes harder to differentiate between the junk and desired states in the measurement and post-processing. When the noise is high enough, no valid information can be extracted, and the model freezes or has no propagation using up to $100$k shots. Despite using the IBM native gates here, the simulation with mid-level noise (which includes the same quantum depolarizing error rates from
Quantinuum H2 and all-to-all connectivity) indicates that useful simulations could be performed with current quantum computers. 
\begin{table}[hbt!]
    \centering
    \caption{Comparison between noiseless QLGA (in red) and different noise levels of the noisy QLGA (NQLGA) (in blue). The number of shots used appears in parentheses. The number of shots is adjusted to the minimum necessary for accurate simulations. Wave propagation is not observed in the model of high noise rate using up to $100$k shots. An ensemble average of fifteen is used. Additionally, the noiseless simulation has been performed with $350$ shots, which is found to measure all the states with high probability for this simulation.}
    \begin{tabular}{@{}c@{}c@{}c@{}c@{}c@{}}
       \toprule
        \diagbox[width=5em]{Noise}{Time}& $t=2$ & $t=16$ & $t=24$ & $t=48$ \\
        \midrule
   {\footnotesize Low ($800$)}
        & \raisebox{-0.5\height}{\includegraphics[width=0.225\linewidth]{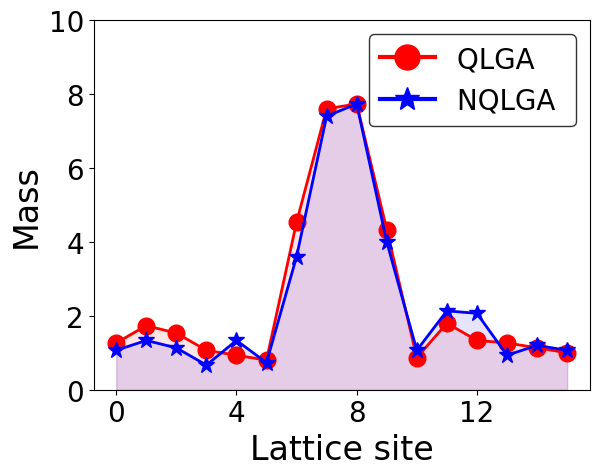}}
        & \raisebox{-0.5\height}{\includegraphics[width=0.225\linewidth]{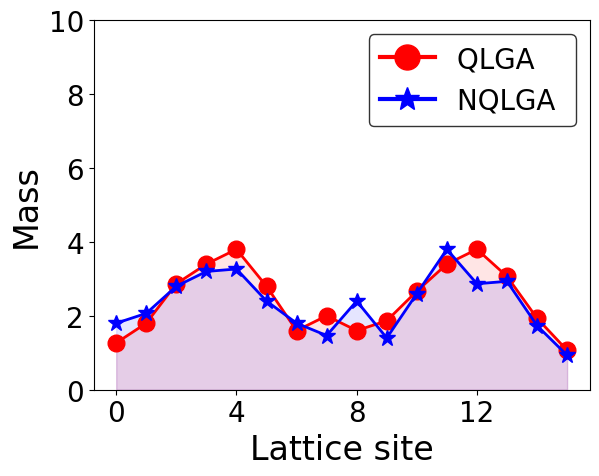}}
        & \raisebox{-0.5\height}{\includegraphics[width=0.225\linewidth]{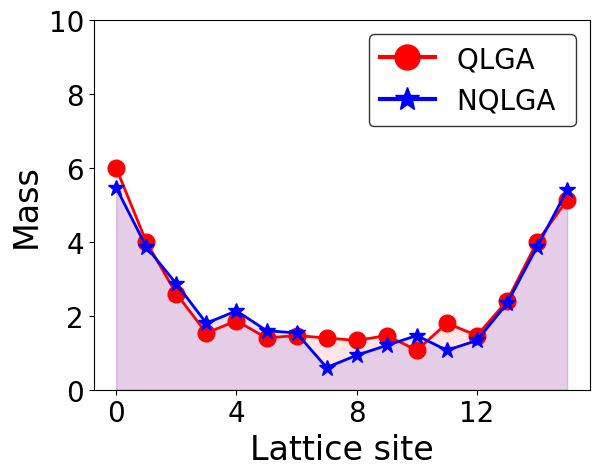}}
        & \raisebox{-0.5\height}{\includegraphics[width=0.225\linewidth]{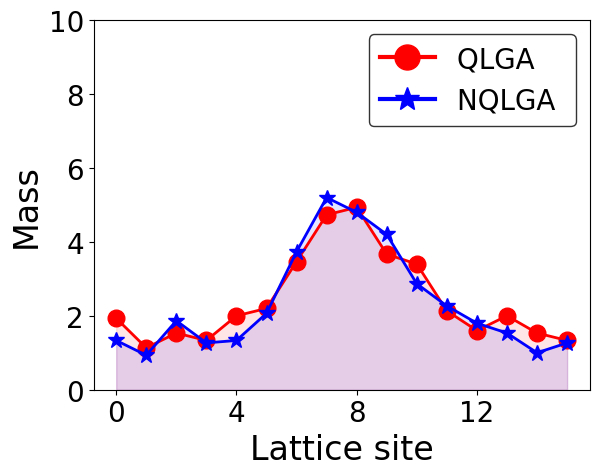}} \\
    {\footnotesize Mid ($3$k)}
        & \raisebox{-0.5\height}{\includegraphics[width=0.225\linewidth]{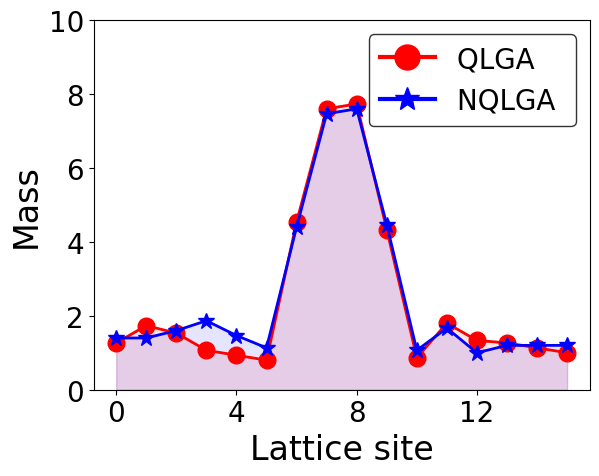}} 
        & \raisebox{-0.5\height}{\includegraphics[width=0.225\linewidth]{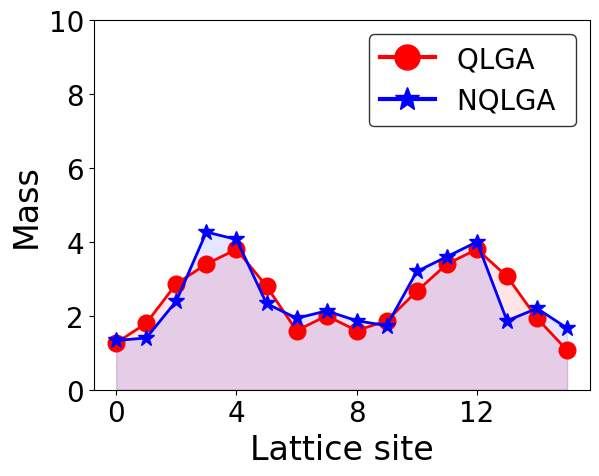}} 
        & \raisebox{-0.5\height}{\includegraphics[width=0.225\linewidth]{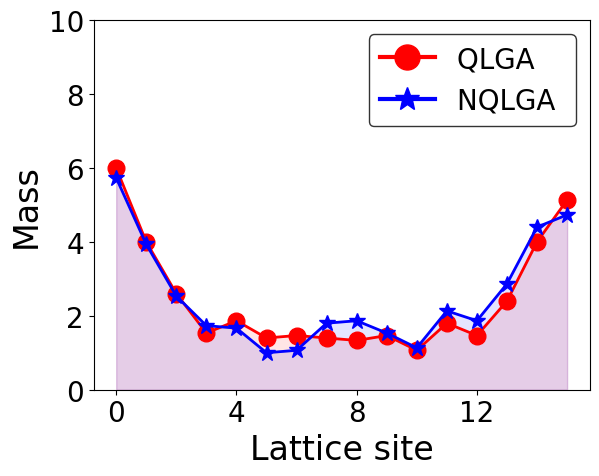}} 
        & \raisebox{-0.5\height}{\includegraphics[width=0.225\linewidth]{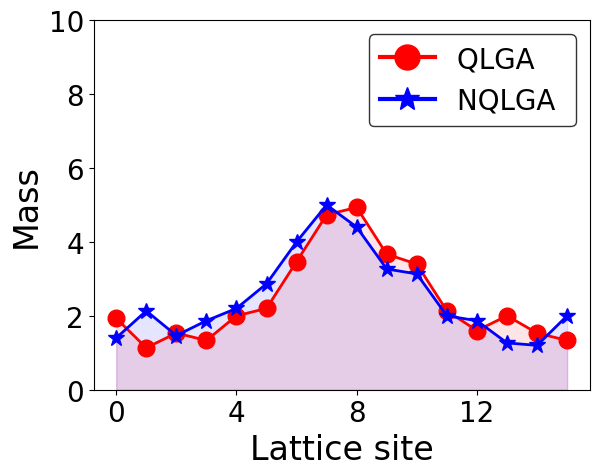}} \\
    {\footnotesize High ($100$k)} 
        & \raisebox{-0.5\height}{\includegraphics[width=0.225\linewidth]{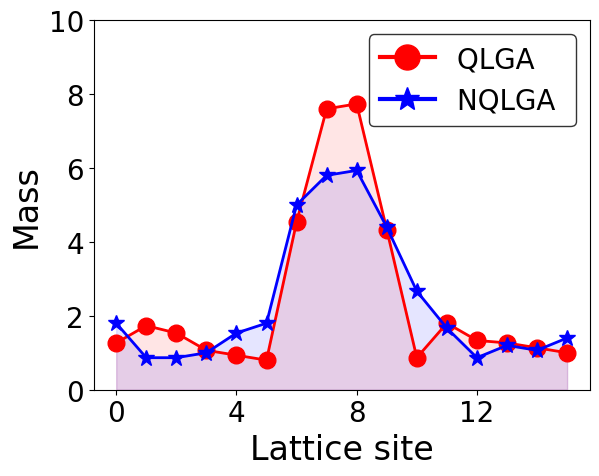}} 
        & \raisebox{-0.5\height}{\includegraphics[width=0.225\linewidth]{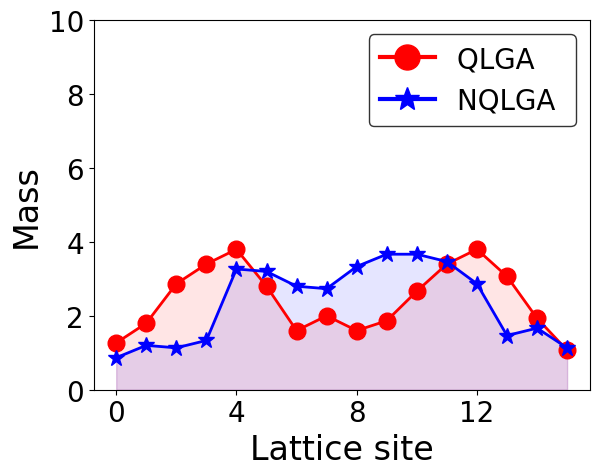}} 
        & \raisebox{-0.5\height}{\includegraphics[width=0.225\linewidth]{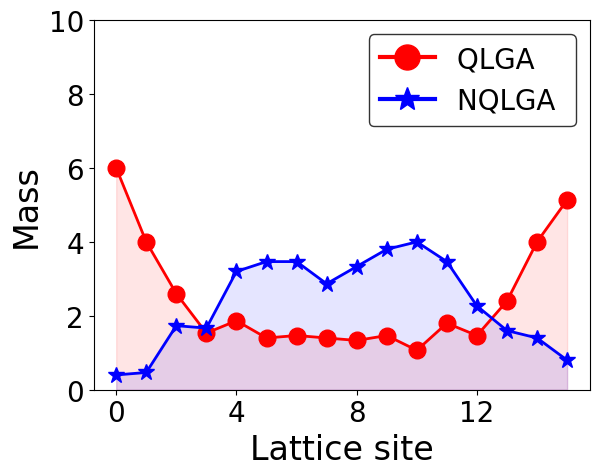}} 
        & \raisebox{-0.5\height}{\includegraphics[width=0.225\linewidth]{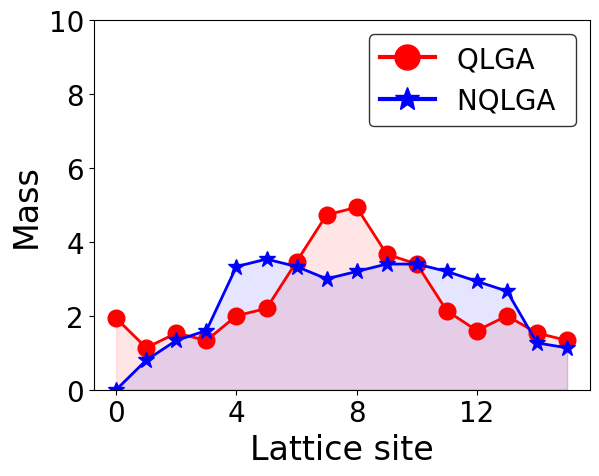}} \\
        \bottomrule
    \end{tabular}
    \label{table:comparison_noise}
\end{table}

We also compare how the simulation progresses under different numbers of shots using a low error rate. As observed in Table~\ref{table:comparison_shots}, the noise slowly decreases with the number of shots. It is not until $600$ shots are used that we begin to differentiate between the junk states and desired states for this specific error rate. However, there is not yet an efficient method for finding the most optimal number of shots for accurate results. Specifically, as seen in the figure, after eight time steps, the information is mostly lost for the case of $200$ shots, and it is difficutl to observe the two characteristic moving waves. The second case with $400$ shots loses part of its information, beginning to have discrepancies during the collision with the wall, which are shown also when the wave collides in the middle once more. Adding further shots does not change the result as there is a high probability of measuring every state up to $48$ time steps. In general, once a particle is not measured correctly, the simulation changes, and additional missing particles build up the previous errors from previous time steps.
This is because only the $n$ most probable states corresponding to the total number of particles are introduced in the next time step during the post-processing. We must notice that the initial condition is created probabilistically for the centered and out regions and small discrepancies are expected. 
\begin{table}[hbt!]
    \centering
    \caption{Low noise QLGA with different numbers of shots represented by $s$. In red: QLGA with 200 shots. In blue: 400 shots. In green: 600 shots.} 
    A spatial average of two sub-lattices 
    and an ensemble average of $15$ are used, therefore we do not see integer mass values. 
    \begin{tabular}{@{}c@{}c@{}c@{}c@{}c@{}}
       \toprule
        $t=1$ & $t=16$ & $t=24$ & $t=48$ \\
        \midrule
        \raisebox{-0.5\height}{\includegraphics[width=0.25\linewidth]{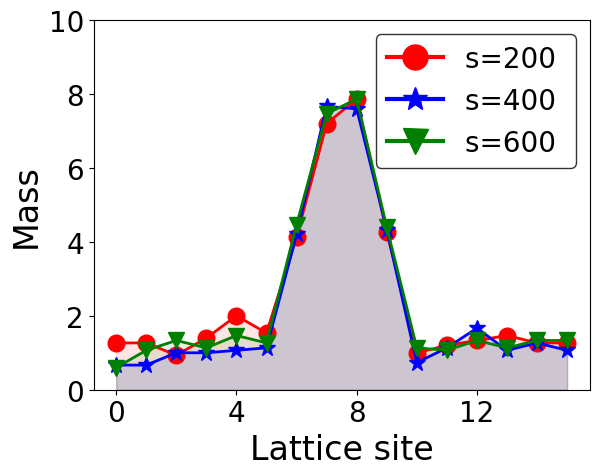}} & \raisebox{-0.5\height}{\includegraphics[width=0.25\linewidth]{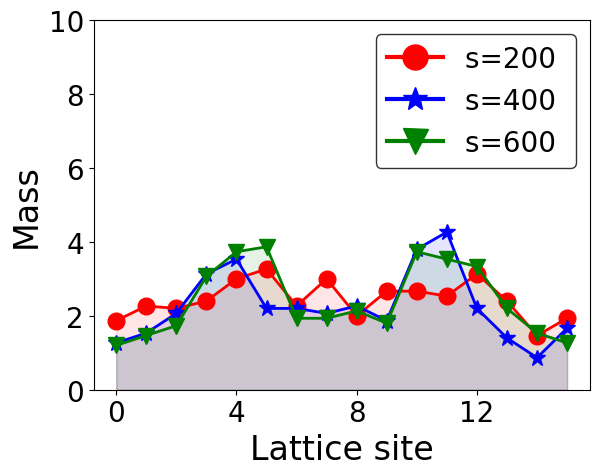}} & \raisebox{-0.5\height}{\includegraphics[width=0.25\linewidth]{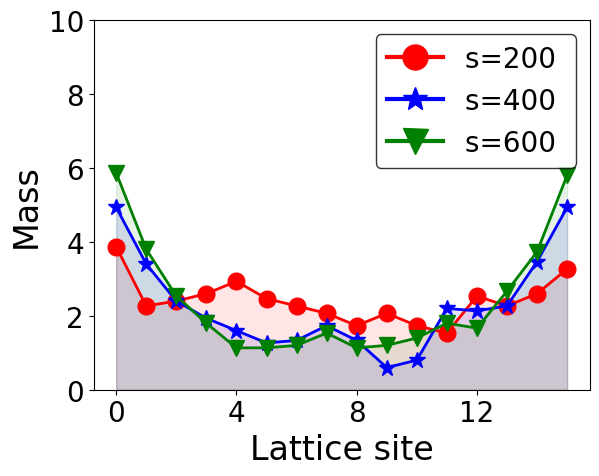}} & \raisebox{-0.5\height}{\includegraphics[width=0.25\linewidth]{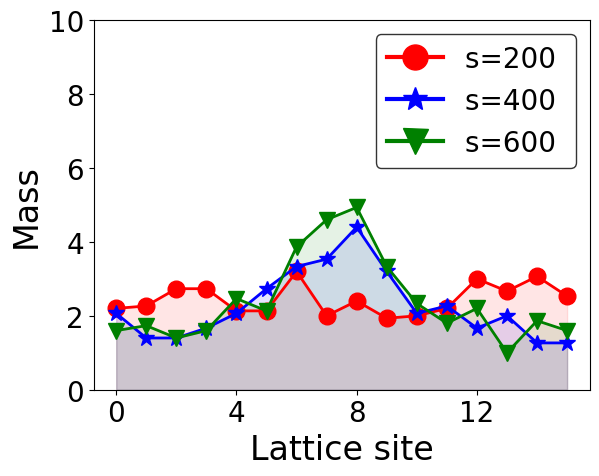}} \\
        \bottomrule
    \end{tabular}
    \label{table:comparison_shots}
\end{table}

\section{Conclusions}
In this work we have presented a novel approach to simulate 1D and 2D lattice gas automata (LGA) with a quantum computer, demonstrating a potential for significant computational advantage in systems with a large number of states. Utilizing the state superposition and parallelization that quantum computing offers, we have designed an algorithm capable of simulating LGA, which has a logarithmic complexity in the number of $CX$ gates with respect to the number of lattice sites. This is in contrast to the classical LGA which has linear complexity with respect to the number of lattice sites~\cite[Ch.~2]{Rivet_Boon_2005}. Furthermore, our study explores the effects of noise on the performance of the algorithm. The results indicate that while high noise levels make distinguishing between the results and computational junk states challenging, accurate simulations could be achievable already with current noisy devices despite the need for increasing shots.

\coloredtext{We found developing a quantum algorithm for multiple time steps to be a significant challenge}, as this necessitates moving from a superposition of channels to a binary encoding. Without a quantum-native time-stepping scheme the algorithm requires a state initialization, which, in its standard form, involves an exponential number of gates and hinders the quantum advantage. However, this is a common challenge in many quantum algorithms, and ongoing research shows promise in overcoming this limitation.

Our work reveals a promising potential for quantum simulations of LGA. \coloredtext{With further developments of a possible algorithm for multiple time steps or an efficient state preparation and noise management}, we anticipate that QLGA will become a valuable tool in computational physics, opening doors to efficient quantum simulation of classical fluid dynamics.
\appendix
\appendixpage
\begin{appendices}{
\section{Extension to 2D: Quantum HPP model}
\label{appendixA}
\subsection{Classical HPP model}

\begin{figure*}[ht]
    \centering
    \usetikzlibrary{arrows.meta}

\tikzset{
  myarrow/.style={
    line width=1.7pt,
  }
}

\begin{tikzpicture}[scale=1, transform shape]

    \draw[->,red,myarrow] (2,-0.05)--(2,-1);
    \draw[->, red,myarrow ] (2,-0.05)--(2,1);
    \draw[->,red, myarrow] (1.95,0)--(1,0);
    \draw[->,red,myarrow] (2.05,0)--(3,0);

    \fill (0,2) circle (2pt);
    \fill (2,2) circle (2pt);
    \fill (4,2) circle (2pt);

    \fill (0,0) circle (2pt) ;
    \fill (2,0) circle (2pt);
    \fill (4,0) circle (2pt) ;

    \fill (0,-2) circle (2pt);
    \fill (2,-2) circle (2pt);
    \fill (4,-2) circle (2pt);

    \draw[line width=0.2mm] (0,2)--(4,2);
    \draw[line width=0.2mm] (0,0)--(4,0);
    \draw[line width=0.2mm] (0,-2)--(4,-2);

    \draw[line width=0.2mm] (0,-2)--(0,2);
    \draw[line width=0.2mm] (2,-2)--(2,2);
    \draw[line width=0.2mm] (4,-2)--(4,2);

	\draw[dotted, line width=0.2mm] (0,-2)--(-2,-2);  \fill (-2,-2) circle(2pt);
	\draw[dotted, line width=0.2mm] (0,0)--(-2,0);  \fill (-2,0) circle(2pt);
	\draw[dotted, line width=0.2mm] (0,2)--(-2,2);  \fill (-2,2) circle(2pt);

	\draw[dotted, line width=0.2mm] (4,-2)--(6,-2);  \fill (6,-2) circle(2pt);
	\draw[dotted, line width=0.2mm] (4,0)--(6,0);  \fill (6,0) circle(2pt);
	\draw[dotted, line width=0.2mm] (4,2)--(6,2);  \fill (6,2) circle(2pt);

 	\draw[dotted, line width=0.2mm] (4,-2)--(4,-4);  \fill (4,-4) circle(2pt);
 	\draw[dotted, line width=0.2mm] (2,-2)--(2,-4);  \fill (2,-4) circle(2pt);
 	\draw[dotted, line width=0.2mm] (0,-2)--(0,-4);  \fill (0,-4) circle(2pt);

   	\draw[dotted, line width=0.2mm] (4,2)--(4,4);  \fill (4,4) circle(2pt);
   	\draw[dotted, line width=0.2mm] (2,2)--(2,4);  \fill (2,4) circle(2pt);
   	\draw[dotted, line width=0.2mm] (0,2)--(0,4);  \fill (0,4) circle(2pt);

\draw (-2,-4)--(0,-4);  \fill (-2,-4) circle(2pt) node[below=1mm] {$(0,0)$}; 
\draw (-2,+4)--(0,+4);  \fill (-2,+4) circle(2pt) node[above=1mm] {$(0,M)$}; 
\draw (4,+4)--(+6,+4);  \fill (+6,+4) circle(2pt) node[above=1mm] {$(N,M)$}; 
\draw (4,-4)--(+6,-4);  \fill (+6,-4) circle(2pt) node[below=1mm] {$(N,0)$}; 

\draw[line width=0.2mm] (-2,-4)--(6,-4);
\draw[line width=0.2mm] (-2,4)--(6,4);
\draw[line width=0.2mm] (-2,-4)--(-2,4);
\draw[line width=0.2mm] (6,-4)--(6,4);

\draw[->,myarrow] (-3.5,-5.5)--(-3.5,-4.5) node[right] {$y$};
\draw[->,myarrow] (-3.5,-5.5)--(-2.5,-5.5) node[below] {$x$};

    \draw  (1.5,-0.11)  node[above=0.1mm] {\large $n_3$};
    \draw  (2.55,0.1)  node[below=0.1mm] {\large $n_1$};
    \draw  (1.9,-0.6)  node[right=0.1mm] {\large $n_4$};
    \draw  (2.1,0.6)  node[left=0.1mm] {\large $n_2$};

\end{tikzpicture}
    \caption{HPP lattice configuration.}
    \label{fig:lattice-hpp}
\end{figure*}
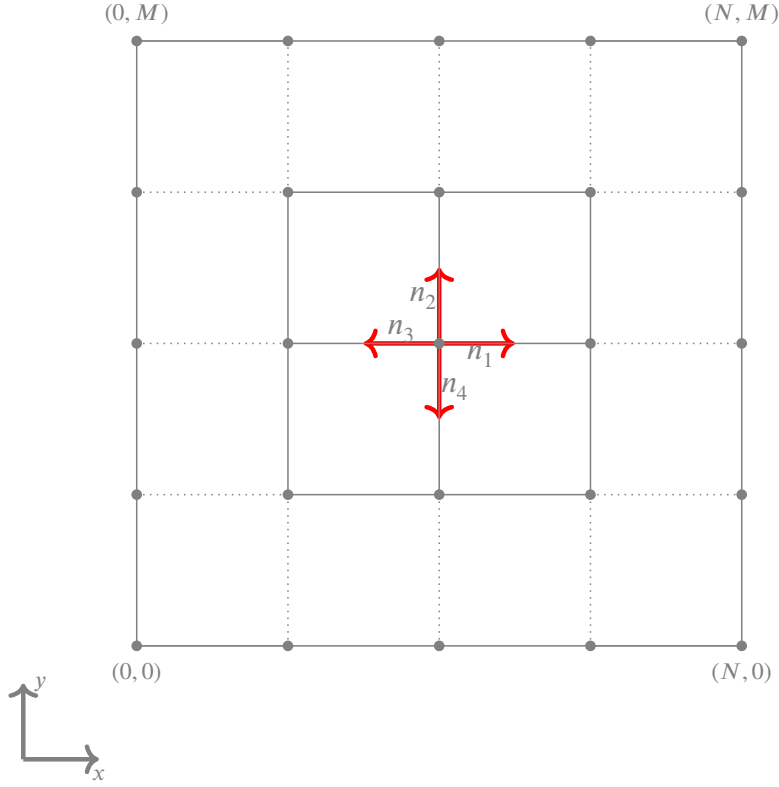

The HPP model is the two-dimensional equivalent of the D1Q3 model, originally developed as a cellular automaton to recreate fluid dynamics in 1970s by Hardy, Pomeau and de Pazzis~\cite{Hardy_Pazzis_Pomeau_1976}. The HPP consists of discrete dynamics of particles moving and colliding on a two-dimensional square lattice, conserving the momentum and the particle number. The model makes use of the von Neumann method, meaning that only the vertical and horizontal neighbors are given four different channels $(n_1 n_2 n_3 n_4)$ (see Figure~\ref{fig:lattice-hpp}).

There are only two simple collision rules. First, if particles going down and up collide on a lattice site as described by the state $(0101)$, the output will be one particle left and another right, giving the state $(1010)$. The opposite applies when a particle going left collides with one going right, creating up-and-down particles. 
The collision operator can then be seen as an identity for all the states except $(1010)\longrightarrow (0101)$ and $(0101)\longrightarrow (1010)$. This property makes the dynamics self-dual and satisfies the principle of detailed balance. During the propagation step, each particle is advanced along the prescribed direction.  

\subsection{Quantum HPP model}
In this section, we expand the algorithm for a 2D HPP case, using the one-dimensional algorithmic structure as a basis, where only the dimensions and the new collision rules change. In particular, we use the binary-based approach. The register $c$ has now four qubits, and each qubit encodes a particular channel. As illustrated in Figure~\ref{fig:lattice-hpp}, a particle with direction $n_i$ is encoded with a corresponding qubit $c_i$. We again use the register $l$ of $n$ qubits to store the position of the lattice sites; as we are now in 2D, we have $n = 2\log_2(N)$ with $N$ being the number of lattice sites along each direction. The role of the register $a$ is to map the channels into the superposition configuration necessary for the propagation step. After initialization we get the state:
\begin{equation}
    \begin{split}
    \qstate{in}=\qop{I}{HPP} \qstate{0} & = \qop{I}{HPP} \left(\ket{0}_l^{\otimes n} \otimes \ket{0}_c^{\otimes 4}\right) \otimes \ket{0}_a^{\otimes 3}  \\ 
        & =  \frac{1}{\sqrt{2^n}}\sum_{i=0}^{2^n-1} d_i \otimes \ket{0}_a^{\otimes 3} =
     \frac{1}{\sqrt{2^n}}\sum_{i=0}^{2^n-1} %
     {\ket{l_1....l_n|c_1 c_2 c_3 c_4}}_{i} %
     \otimes \ket{0}_{a1} \otimes \ket{0}_{a2} \otimes \ket{0}_{a3}.
    \end{split}
\end{equation}
The $x$ and $y$ coordinate indices of the lattice sites are mapped into a one-dimensional index array using the relation $k=y + xN$, with $x,y \in \{0,\ldots,N-1\}$, therefore giving $k \in \{0,\ldots,N^2-1\}$ as desired. Note that we are using a square lattice configuration domain with $N=M$ as in Figure~\ref{fig:lattice-hpp}. As with the binary-encoded D1Q3 model, this encoding is \emph{ideal} in that we desire to have exactly one channel state per lattice site instead of a superposition of states.

The collision step affects only two particular states as explained in Section~\ref{sect:Chapter 2}. This results in the following state:
\begin{equation}
  \qstate{col}
    = \qop{C}{HPP} \qstate{in} 
    = \frac{1}{\sqrt{2^n}}\sum_i \ket{l_1....l_n~|~c'_1 c'_2 c'_3 c'_4}_i \otimes \ket{0}_a^{\otimes 3} 
    = \frac{1}{\sqrt{2^n}}\sum_i \ket{l_1....l_n~|~\sigma(c_1 c_2 c_3 c_4)}_i \otimes \ket{0}_a^{\otimes 3} ,
\end{equation}
where the channel map $\sigma: C \to C$ is defined as
\begin{equation}
  \sigma(c_1 c_2 c_3 c_4) = 
    \begin{cases}
      1010 , & ~\text{for}~ \, c_1 c_2 c_3 c_4 = 0101 \\
      0101 , & ~\text{for}~ \, c_1 c_2 c_3 c_4 = 1010 \\
      c_1 c_2 c_3 c_4 , & ~\text{otherwise}.\ 
    \end{cases}
\end{equation}

Following the collision, we map the result into the superposition of lattice sites and channels in the same manner as in the case of D1Q3 QLGA. The only difference is that now we use the full superposition of the qubits $a_2$ and $a_3$ to store the four HPP channels. The state after the mapping $\qop{M}{HPP}$ operator is    
\begin{equation}
    \begin{split}
    \qstate{map} &= \qop{M}{HPP} \qstate{col} \\
    & = \mathsc{MCSWAP}_{a-c} 
     \frac{1}{\sqrt{2^n}}\sum_i \ket{l_1....l_n|c'_1 c'_2 c'_3 c'_4}_i \otimes \ket{0}_{a1} \otimes H \ket{0}_{a2} \otimes H \ket{0}_{a3} \\ 
     & = \frac{1}{2\sqrt{2^n}} \left( \sum_i \ket{l_1....l_n|c'_1 c'_2 c'_3 a_1}_i \otimes \ket{c'_4} \otimes \ket{00}_{a2,a3} \right.\\  
     & + \sum_i \ket{l_1....l_n|c'_1 a_1 c'_3 c'_4}_i \otimes \ket{c'_2} \otimes \ket{10}_{a2,a3} \\ 
     & + \sum_i \ket{l_1....l_n|a_1 c'_2 c'_3 c'_4}_i \otimes \ket{c'_1} \otimes \ket{01}_{a2,a3} \\ 
     & + \left. \sum_i \ket{l_1....l_n|c'_1 c'_2 a_1 c'_4}_i \otimes \ket{c'_3} \otimes \ket{11}_{a2,a3} \right).
    \end{split}
\end{equation}
The quantum circuit for both the collision and mapping steps is shown in Figure~\ref{fig:init_coll_hpp}. 
\begin{figure}
    \centering
    \input{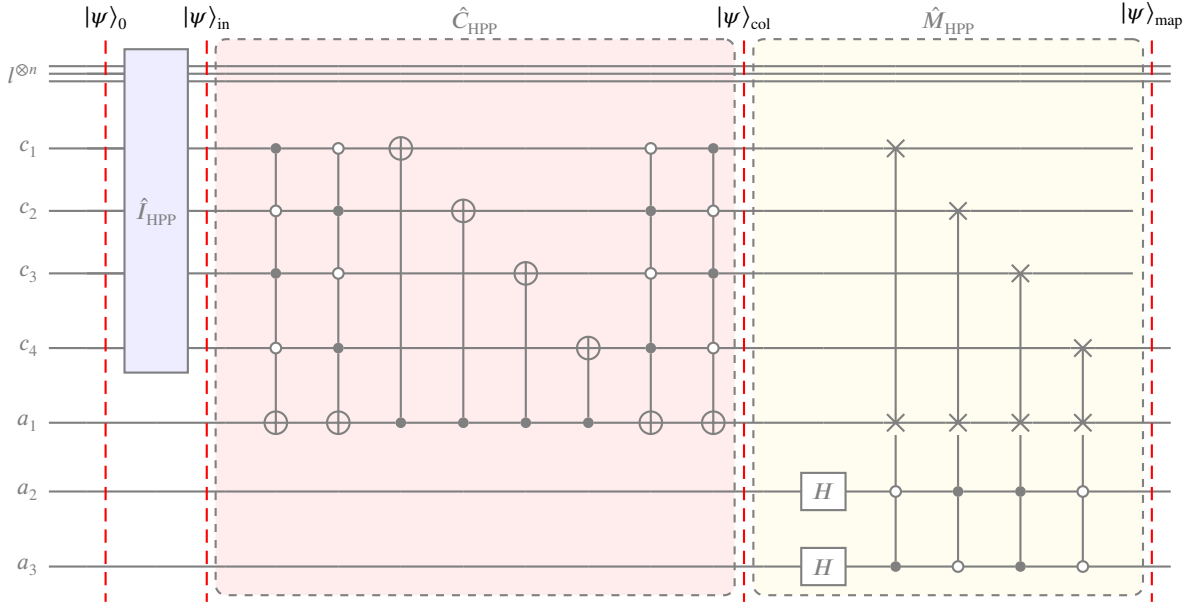}
    \caption{Quantum circuit for the initialization and collision step -- the HPP model.}
    \label{fig:init_coll_hpp}
\end{figure}

Propagation in four different orthogonal directions on a two-dimensional lattice can be constructed by sandwiching a quantum circuit for the left/right propagation (such as in Figure~\ref{fig:prop}) between a set of controlled swap operators. This gives a permutation of the substates for the particles propagating in the left and right directions, effectively parallelizing the four directions. An example of a quantum circuit for the propagation on $16\times{}16$ lattice (corresponding to eight qubits in the register $l$) is shown in Figure~\ref{fig:prop_hpp}. After the propagation step we obtain a new state $\qstate{prop}$:
\begin{equation}
    \begin{split}
    \qstate{prop}=\qop{P}{HPP} \qstate{map} & = \qop{P}{HPP} \frac{1}{2 \sqrt{2}} \left( \sum_i \ket{l_1....l_n|c'_1 c'_2 c'_3 a_1}_i \otimes \ket{c'_4} \otimes \ket{00}_{a2,a3} \right.\\  
     & + \sum_i \ket{l_1....l_n|c'_1 a_1 c'_3 c'_4}_i \otimes \ket{c'_2} \otimes \ket{10}_{a2,a3} \\ 
     & + \sum_i \ket{l_1....l_n|a_1 c'_2 c'_3 c'_4}_i \otimes \ket{c'_1} \otimes \ket{01}_{a2,a3} \\ 
     & + \left. \sum_i \ket{l_1....l_n|c'_1 c'_2 a_1 c'_4}_i \otimes \ket{c'_3} \otimes \ket{11}_{a2,a3} \right)\\
     & = \frac{1}{2 \sqrt{2}} \left( \sum_i \ket{l_1....l_n}_i \otimes \ket{c'_1 c'_2 c'_3 a_1}_{i+1} \otimes \ket{c'_4} \otimes \ket{00}_{a2,a3} \right.\\  
     & + \sum_i \ket{l_1....l_n}_i \otimes \ket{c'_1 a_1 c'_3 c'_4}_{i-1} \otimes \ket{c'_2} \otimes \ket{10}_{a2,a3} \\ 
     & + \sum_i \ket{l_1....l_n}_i \otimes \ket{a_1 c'_2 c'_3 c'_4}_{i+p} \otimes \ket{c'_1} \otimes \ket{01}_{a2,a3} \\ 
     & + \left. \sum_i \ket{l_1....l_n}_i \otimes \ket{c'_1 c'_2 a_1 c'_4}_{i-p} \otimes \ket{c'_3} \otimes \ket{11}_{a2,a3} \right),
    \end{split}
\end{equation}
where $p = N$ with $N$ being the number of lattice sites along one dimension (Figure~\ref{fig:lattice-hpp}). Again, we assume periodicity on the lattice, but in the 2D case the periodic indexing has to be interpreted with respect to both axes, leading to a slight abuse of notation in the equation above.
\begin{figure}[ht]
    \centering
    \input{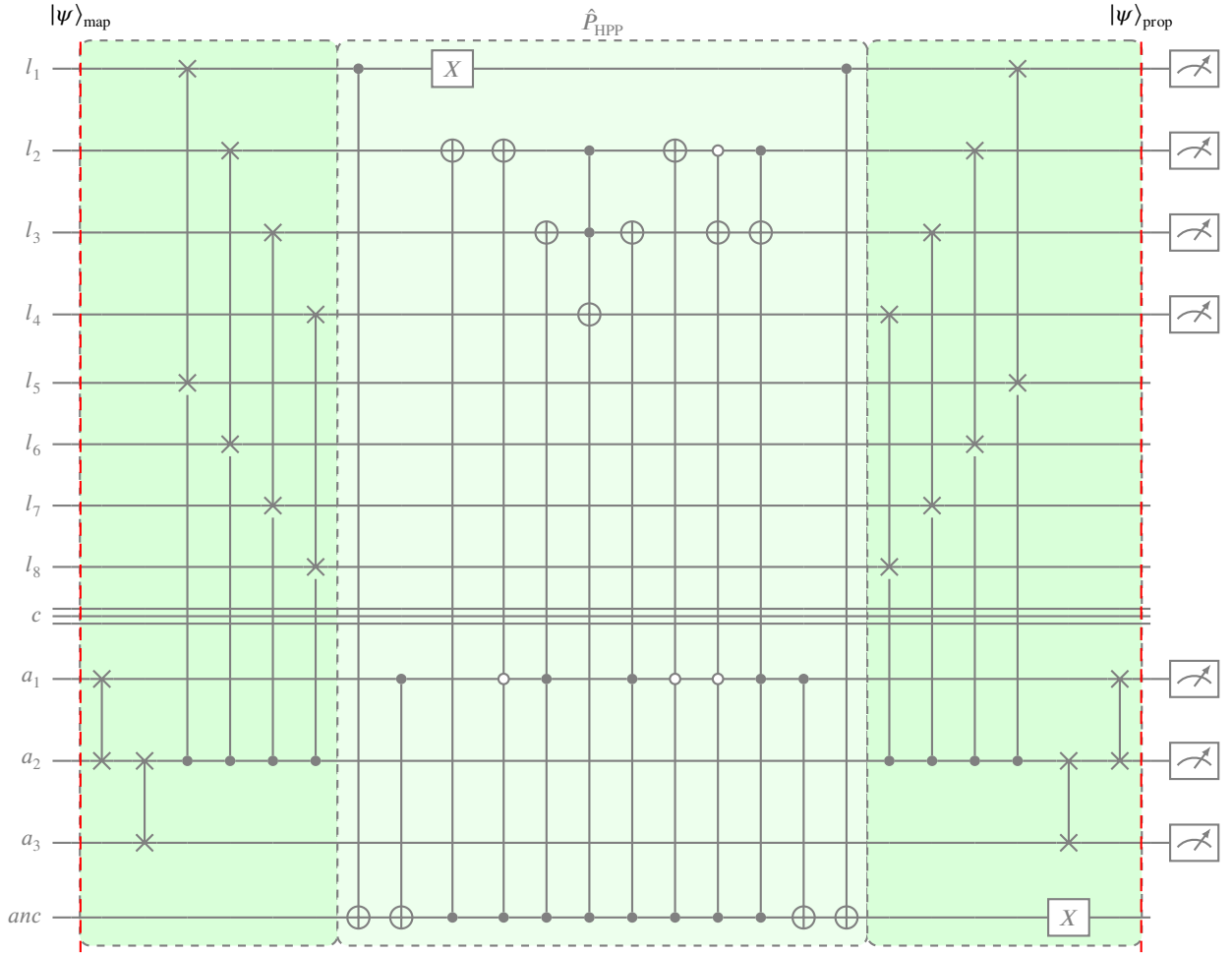}
    \caption{Quantum circuit for the propagation step in the HPP model.}
    \label{fig:prop_hpp}
\end{figure}

}
\end{appendices}
\printcredits

\section*{Role of the funding source}

This research was supported by the Business Finland project 9820/31/2022 Quantum-Native Multiphysics.

\section*{Declaration of competing interest}
The authors declare no competing interests.

\bibliographystyle{model1-num-names}

\bibliography{references}

\end{document}